\documentclass[11pt,a4paper]{article}

\usepackage{adjustbox}
\usepackage{amsfonts}
\usepackage{amssymb}
\usepackage{amsmath}
\usepackage{amsthm}
\usepackage{authblk}
\usepackage{array}
\usepackage{bbm}
\usepackage{booktabs}
\usepackage{caption}
\usepackage{subcaption}
\usepackage{dsfont}
\usepackage{enumerate}
\usepackage{framed}
\usepackage{graphics}
\usepackage{graphicx}
\usepackage{hyperref}
\usepackage{inputenc}
\usepackage{latexsym}
\usepackage{mathrsfs}
\usepackage[table]{xcolor}
\usepackage{tikz}
\usepackage[normalem]{ulem}
\usepackage{makecell}
\usepackage{lmodern,babel,adjustbox,booktabs,multirow}
\usepackage{multicol}
\usepackage[style=english]{csquotes}

\title{SHREC 2022: Protein-ligand binding site recognition}

\author[1,\thanks{Track organizer}]{Luca Gagliardi}
\author[2,$^*$]{Andrea Raffo}
\author[2,$^*$,\thanks{Corresponding author}]{Ulderico Fugacci}
\author[2,$^*$]{Silvia Biasotti}
\author[1,$^*$]{Walter Rocchia}
\author[3]{Hao Huang}
\author[4]{Boulbaba Ben Amor}
\author[5]{Yi Fang}
\author[6]{Yuanyuan Zhang}
\author[6]{Xiao Wang}
\author[6]{Charles Christoffer}
\author[7]{Daisuke Kihara}
\author[8]{Apostolos Axenopoulos}
\author[8]{Stelios Mylonas}
\author[8]{Petros Daras}

\affil[1]{CONCEPT Lab, Istituto Italiano di Tecnologia (IIT), Via Enrico Melen 83, 16152 Genova, Italy}
\affil[2]{Istituto di Matematica Applicata e Tecnologie Informatiche  ``E. Magenes'', Consiglio Nazionale delle Ricerche, Via de Marini 6, 16149 Genova, Italy}
\affil[3]{Computer Science and Engineering, Tandon School of Engineering, New York University, 6 MetroTech Center, Brooklyn, New York, USA}
\affil[4]{Inception Institute of Artificial Intelligence, Masdar City, Abu Dhabi, UAE}
\affil[5]{Electrical and Computer Engineering, New York University Abu Dhabi \& Tandon School of Engineering, New York University, Saadiyat Island, Abu Dhabi, UAE}
\affil[6]{Department of Computer Science, Purdue University, 249 S. Martin Jischke Dr. West Lafayette, IN 47907, USA}
\affil[7]{Department of Biological Sciences, Department of Computer Science, Purdue University, 249 S. Martin Jischke Dr. West Lafayette, IN 47907, USA}
\affil[8]{Information Technologies Institute, Centre for Research and Technology Hellas, 57001, Greece}

\date{}                     
\begin{document}
\maketitle

\begin{abstract}
This paper presents the methods that have participated in the SHREC 2022
contest on protein-ligand binding site recognition. The prediction of protein-
ligand binding regions is an active research domain in computational biophysics
and structural biology and plays a relevant role for molecular docking and
drug design. The goal of the contest is to assess the effectiveness of
computational methods in recognizing ligand binding sites in a protein based
on its geometrical structure. Performances of the segmentation
algorithms are analyzed according to two evaluation scores
describing the capacity of a putative pocket to contact a ligand and to pinpoint the correct binding region.
Despite some methods perform remarkably, we show that simple non-machine-learning approaches remain very competitive against data-driven algorithms.  In general, the task of pocket detection remains a challenging learning problem which suffers of intrinsic difficulties due to the lack of negative examples (data imbalance problem).
\end{abstract}

\section{Introduction}

The general objective of this SHREC track is to evaluate the effectiveness of computational methods in recognizing most likely protein-ligand binding sites based on the geometrical structure of the protein. Starting from a set of protein-ligand complex structures obtained via X-ray crystallography and deposited in the PDB repository, we build the proteins’ Solvent Excluded Surface (SES) \cite{Connolly} via the freely available software NanoShaper (NS) \cite{DeCherchi2013,Decherchi2019}. Additionally, for each structure we provide an anonymized PQR file (neither residues, nor atomic names and charges) containing the atomic centers and radii.
The track is jointly organized by IMATI-CNR and the CONCEPT Lab at IIT.

\paragraph{Motivation}
The molecular surface of a protein, often defined as the separating surface between solvent (water) accessible and inaccessible regions \cite{LeeRichards,Connolly,Chen2010}, plays a fundamental role in the characterization and prediction of the interactions of a protein with other biomolecules. In this study, we aim to identify pockets able to bind ligands, based on their shape. The prediction of protein-ligand binding region is one of the focal points of activity in computational biophysics and structural biology. Indeed, when a small molecule binds to a protein, it affects its biological behavior. Therefore, the identification of candidate binding sites is a key aspect, which is essential and preparatory to drug design. Adequate computational techniques able to predict these regions of potential interaction are important because they can leverage the information obtained from the growing number of known (co-crystallized) protein-ligands systems and lead to innovative therapeutic strategies.

\paragraph{Description of the track}
A dataset of approximately 1090 protein surfaces and corresponding to about 1720 relevant binding sites (regions close to the ligand) is provided to the participants. The dataset is split into a training set and a test set (in the proportion 85-15).

\begin{figure}[htb!]
    \begin{center}
    \begin{tabular}{cc}
        \includegraphics[scale=0.2]{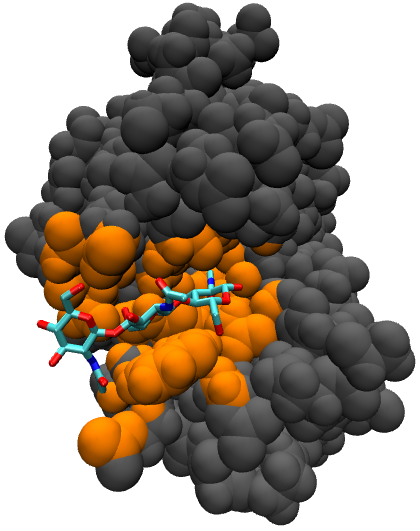}&
        \includegraphics[scale=0.5]{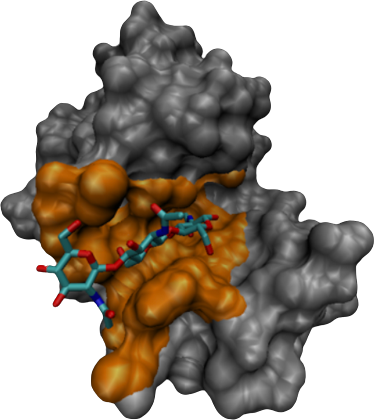}
       \\ & \\
        (a) & (b)
    \end{tabular}
    \end{center}
    \caption{Hen Egg White Lysozyme (PDB code, 1hew) a) Atom spheres representation (PQR). b) SES realized with NanoShaper. The binding site of the depicted ligand is highlighted in orange with: a) atoms within 5 \AA from the ligand, b) surface vertices within 4 \AA from the ligand.}\label{fig:dataset}
\end{figure}
For the training set we identify the set of vertices which are within 4 \AA of any ligand atom center, and we label in the PQR the set of atoms whose center is within 5 \AA of any ligand atom center. We hereby refer to these sets of vertices and atoms as \emph{binding pockets}.
An example of protein surface and of the corresponding ligand binding site is depicted in Figure \ref{fig:dataset}.
The information regarding ligand binding sites is provided in a separate TXT file.
The TXT file represents the ligand binding sites of the cognate structure via a vector containing, for each vertex of the corresponding OFF file (and in the same order), either zero (i.e., this vertex is not known to contribute to a binding site) or a strictly positive integer (different values code for distinct protein-ligand binding sites). The same information is replicated in the one-but-last (i.e., the charge) column of the PQR file.

To compare the performance of the candidate methods, we ask the participants to provide us with a vector representing the 10 most likely binding sites they identify for each protein in the test set, either in terms of the vertices (if using the OFF files) or of the atoms (if using the PQR files). We further ask the participants to provide a ranking of them, from the most to the least likely.
We want to highlight that:
\begin{enumerate}
    \item A single structure can contain more than one binding site (more than one co-crystallized ligand).
    \item The training set does not imply any ranking. All provided binding pockets are positive examples and should be considered equally important.
\end{enumerate}
The remaining of this paper is organized as follows. In Section \ref{sec:star} we offer a perspective on the state of the art and point to previous benchmarks/reviews that aim at comparing pocket identification algorithms.
We also discuss the general challenges associated to protein binding site learning and prediction.
Then, in Section \ref{sec:dataset} we detail the dataset and its post-processing, and in Section \ref{sec:eval_meas} the classification metrics used in the contest. The methods submitted for evaluation to this SHREC are detailed in Section \ref{sec:methods}, while their outcomes are presented in Section \ref{sec:results}. Finally, discussions and concluding remarks are in Section \ref{sec:obs_conlusion}.

\section{State of the art and related benchmarks}
\label{sec:star}

Given the molecular surface of a protein, in general a pocket is a concave region of the molecular surface which is accessible from the solvent (e.g., clefts/grooves or invaginations).

The detection of pockets and cavities has been a long-standing challenge in the biophysical community. Therefore a large amount of algorithms have been proposed to tackle this problem.
Such algorithms can be partitioned into three broad categories \cite{Simoes2017}:
a) Evolutionary and template based algorithms (based on multiple sequence alignments to find the location of binding sites on a given protein). These algorithm are mainly addressing the problem from the chemical/biological perspective.
b) Energy-based algorithms (binding sites are detected by computing the interaction energies between protein atoms and a small-molecule probe). These algorithms tackle the problem from a physical prospect. c) Geometric algorithms, which focus on the geometric properties of the molecular surface to detect cavities that may be binding sites.
The more recent tendency is to use a mixture of the three approaches to exploit at maximum the available information.
We focus here on geometric methods. These can be roughly partitioned into four main sub-categories:
1. Sphere based methods: pockets and cavities are defined by filling voids with probe spheres.
2. Grid based methods: a three-dimensional grid scanning is performed which establishes for each grid point whether it belongs or not to the protein. Then, according to different geometric criteria, the grid points belonging to pockets/cavities are identified.
3. Tessellation/alpha shape methods: they rely on the generation of the Delaunay triangulation of the molecular surface and the definition of pockets via filtered sub-complexes of this triangulation (alpha-shape) which define voids on the surface \cite{Edelsbrunner1998}.
4. Surface-based methods: a much smaller set of methods focusing on binding site identification via local analytical geometric properties of the molecular surface (e.g., curvature).

Whatever the pocket generation is, the final step involves the ranking of the putative binding pockets which have been identified.
Several methods return a ranking which is based on an predetermined scoring which is not trained. Despite its simplicity, one of the most widely adopted and successful scoring systems is to use volume ranking \cite{Macari2019}. Some algorithms such as grid based methods, also consider algorithm-specific descriptors (such as the degree of buriedness \cite{Marchand2021}) or chemical information based on the degree of evolutionary conservation of each residue \cite{Huang2006,Glaser2004}.

Another approach is to rank putative pockets according to more complex descriptors trained over datasets by fitting a numerical function via some regression technique \cite{LeGuilloux2009,Halgren2009,Hajduk2005}.
For instance, the Fpocket \cite{LeGuilloux2009} algorithm (an open source method based on Voronoi tessellation available at \cite{fpocket}, which is widely used in the literature) employs five descriptors and returns a ligand coverage score as an independent variable. The descriptors are five independent values that depend on: i) number of alpha spheres, ii) local mean hydrophobic density, iii) proportion of apolar alpha sphere, iv) a polarity score defined as a binarity sum of the polarity over the amino acids in a pocket, v) the alpha sphere density, i.e., the average of the pair to pair distances among the alpha spheres in a pocket.
The Fpocket standard parameters are determined using semi combinatorial/empirical optimisation procedure trained on 307 structures.
Another approach which has been explored is to learn a (druggability) score by machine learning methods, such as,  logistic regression \cite{Schmidtke2010} and Support Vector Machine \cite{Volkamer2012}. This score reflects the probability of a binding site to be druggable.
A drawback of this type of approaches is that a choice must be made upon a (small) set of parameters on which to build the regression. This requires that a good degree of a priori knowledge is available to select the most relevant parameters, which might considerably vary according to the considered system and problem. On the other hand, if one considers a large set of (probably redundant) variables, the risk of overfitting and of losing generality, further increases. Another potential problem is to rely too much on the fitting over the chosen score. For instance this could be a ligand coverage score, representing the fraction of ligand in contact with the binding site (see Section \ref{sec:eval_meas}) \cite{LeGuilloux2009}, or the hit rate of a binding site inferred through a large screening database \cite{Hajduk2005}.
A nice exception among linear combination of variables used to obtain a scoring function for druggability is the one proposed by Cheng et al. \cite{Cheng2007}, where a model-based approach yields a very good prediction of druggability
using solely fundamental biophysical principles based on surface features of the target binding site (curvature and binding site surface area) and some pre-determined constants.

More recently, a new family of data driven approaches has been proposed, which employs a reverse strategy. Rather than using ML to classify and rank previously generated putative pockets, they focus on predicting high-binding-probability small areas as such and then grouping them via an unsupervised clustering approach so as to construct a binding pocket \cite{Macari2019,Krivak2018}.
An advantage of the latter kind of methods is that they are able to return a rather small set of putative pockets compared to the more standard approaches since the selection of valid binding surface points happens upstream, before the actual pocket generation.
Methods falling in this category, heavily rely on chemical information which is gathered (learned) from large protein-ligand binding datasets containing labelled examples \cite{Schmidtke2010,Fauman2011,Volkamer2012}, which often rely on some (strong) assumptions on the negative examples (e.g.,  regions not observed in contact with the ligand, are labelled as non-binding) \cite{Jimenez2017}.

Very accurate comparative reviews can be found in the literature that discuss pros and cons of each method both in terms of outcome and performance \cite{Chen2011,Simoes2017,Krivak2018,Macari2019}.

In this work we aim at assessing and supporting approaches which can be of interest for the computer graphics community, hence those based on geometric and ML-driven techniques. This is why we focus on methods which use geometric properties of molecular surfaces (surface-based methods) in combination or driven by ML approaches.
These types of approaches, and especially the combination with ML (without any use of chemical information), are still relatively poorly explored for the pocket detection task.

This SHREC contest differs from previous SHREC contests related to proteins retrieval and classification, e.g.,  \cite{LANGENFELD2020,RAFFO2021}, because the focus here is the identification of delimited binding sites rather than the comparison of the whole molecular surface or its domains. Moreover, it also differs from contests on the classification of cryo-electron tomograms, e.g., \cite{GUBINS2020}, because the structures we consider are obtained at a finer level of resolution, and we are not focusing in the interaction of a complex system of thousands of proteins.

\subsection{Discussion}
\label{sec:overview}

Learning the inherent features of a ``druggable'' pocket is a very problematic task. Indeed, already the construction of a balanced training set for conventional supervised learning is an ill-posed problem  
since one is forced to start from 3D structures obtained via x-ray crystallography. As a matter of fact, these experiments point to positive samples, but there is no practical way to identify negative labelled ones.
Indeed, an empty pocket does not imply the pocket to be undruggable \cite{Amaro2019}. That is why we can claim that we have available only positive examples in our dataset, as proposed in the training set of this Shrek track.
Furthermore, the ligand binding process may cause structural rearrangements of the protein around the binding region, further complicating the task of identifying promising candidate sites in the so-called ``apo'' structure of the protein \cite{Surade2012}.

The problem of pocket retrieval thus appears as an instance of a \emph{one-class discrimination} problem \cite{Aguti2022}.
One-class discrimination is a learning task that typically arises in
outlier (anomaly) detection or, more generally, in binary discrimination data
mining problems where obtaining examples of one class can be too
expensive or daunting, or where examples of one class are largely under represented (data imbalance) \cite{Itani2020,Decherchi2017}.
Different approaches are used in the literature to solve one-class or data imbalance problems. These are mainly based on two strategies. 1. Fit a probability distribution whose support includes as much as possible the positive data \cite{Decherchi2017} (anomaly detection approaches such as SVDD \cite{Jiang2019}, or Isolation Forest \cite{Liu2008})\footnote{This does not imply that the distribution is always explicitly fitted, but the concept of hyperplanes or hyperspheres containing positive instances is always embedded.}. 2. Use standard binary classificators by generating meaningful instances of the negative class (by some a priori assumptions on the distribution of the negative examples/outliers generation \cite{Itani2020}).

\section{Dataset and data preparation}
\label{sec:dataset}

In this work, we extract an original set of protein-ligand complexes from the binding-MOAD database \cite{Benson2007}.
The set of structures is selected by considering complexes with ligand molecular weights larger than 200 Da, a resolution better than 2\AA, binding data available, and removing redundant structures ($>90\%$ sequence identity).
The Binding MOAD database enables the separation of biologically-relevant ``valid'' ligands from ``invalid'' ones, which have no biological function.

The database is processed using a custom, freely distributed, Python script \cite{lFetch_github}  which: i) determines which subset of the PDB file represents the valid ligand according to the binding MOAD website information; ii) removes the ligand and Heterogenous atoms (HETATM) from the input PDB file and creates a PQR file using the AMBER force field via the pdb2pqr software \cite{Unni2011}; iii) exports the valid ligands heavy atoms in a xyz file; iv) creates a text file containing a map between structure and ligand(s) discarding any invalid ligand and any ligand which has no full correspondence with what expected from the MOAD website naming scheme.
Furthermore, we also excluded from the database the (very large) structures containing more than 10000 lines in the PQR format.
This results in an initial database of 1100 structures and 1808 ligand binding sites.
Then, we assembled the information used in this track. First, using NanoShaper, from the original PQR file we build the triangulation (in OFF format) of the SES molecular surface. Given the (valid) ligands heavy atom coordinates, the ground truth binding region is determined by measuring the distances between ligand atoms and triangulation vertices or protein atoms: vertices within 4 \AA from any ligand atom center are flagged in a separate TXT file where each entry is associated to a line in the OFF file which is either zero (not known to contribute to any binding site) or a positive integer, where different numbers identify distinct protein-ligand binding sites;
protein atom centers within 5 \AA are flagged in the same way by substituting the charge column of the PQR file (one-but-last column).
The PQR file is then anonymized by substituting all atom names with ``C'', residues with ``UNK'', and randomly shuffling all lines.
In a post-processing step, we check also for redundant binding sites in those structures containing multiple co-crystallized ligands.
When a structure contains more than one co-crystallized ligand, we established the amount of overlap between different binding regions and discard regions which are very overlapping. Considering as binding region the protein atoms close to the ligand as defined above (flagged lines of the PQR), we consider two measures: the number of protein atoms which lie in the intersection of the protein-ligand binding site normalized by the number of atoms of each site, and the Jaccard index\footnote{Given two sets $\mathcal{A}$, $\mathcal{B}$, the Jaccard index is
\begin{equation*}
		J = \frac{|\mathcal{A}\cap \mathcal{B}|}{|\mathcal{A}\cup \mathcal{B}|}.
\end{equation*}
}.
If the Jaccard index is larger or equal than $30\%$ or if any of the normalized intersections is larger or equal than $40\%$, we exclude the most overlapping binding site (the one with highest normalized intersection). This results in a set of 1750 well defined ligand binding sites.

Finally, a last post-processing on the flagged vertex is performed, resulting in 1091 structures and 1721 binding sites. More precisely, each graph behind the mesh data structure is analyzed to: discard molecular surfaces with more than one geometric connected component, whenever such additional components are not inner cavities; clean the segments represented by the flagged vertices to get rid of undesired holes and possible secondary (connected) components in the texture.
This set is split into a training and a test set in the proportion 85-15. All flags are removed from the test set before handling the data to the participants.

\section{Evaluation measures}
\label{sec:eval_meas}

In the following, any co-crystallized ligand is reduced to its subset of heavy atoms lying within $5$ \AA of any protein atom. This avoids considering in the evaluation parts of the ligands which protrude into the solvent.

\begin{figure}[htb!]
\centering
  \includegraphics[scale=0.28]{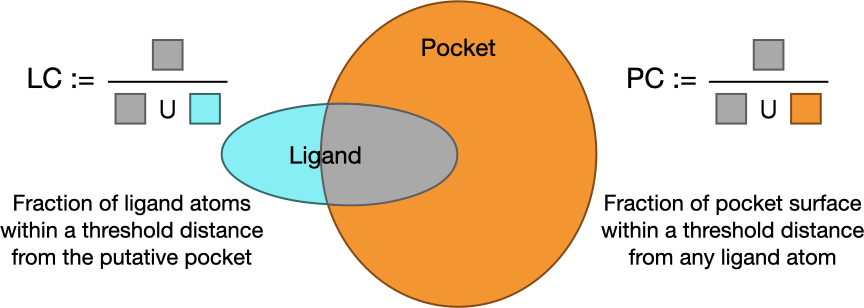}
  \caption{A visual representation of LC and PC scores. Reworking of a picture in \cite{Volkamer2010}.}\label{fig:evaluation}
\end{figure}

Inspired by state-of-the-art biophysical pocket detection methods  \cite{LeGuilloux2009,Volkamer2010}, we here adopt a figure of merit based on the combination of two scores.
\paragraph{Ligand Coverage Score} It represents the fraction of ligand heavy atoms (i.e., excluding Hydrogen) within a threshold distance of the protein atoms (PQR file) or of the surface vertices (OFF file) that compose a putative pocket.
For a given pocket, indicating with $d(i,j)$ the Euclidean distance between entries $i$ and $j$, which can be either atom centers or vertices composing the pocket set  $\mathcal{P}$, and ligand(s) heavy atom centers, which form the $\mathcal{L}$ set, we have
\begin{equation*}
	\mathrm{LC} = \frac{1}{n_{L}}\sum_{j=1}^{n_L}\delta_{ij}
	\quad\text{for }\forall\, i\in \mathcal{P}\,\,\text{with }\delta_{ij} =
	\begin{cases}
		1\,\,\text{if } d(i,j)\leq d^*\\
		0 \,\,\text{if } d(i,j)> d^* 
	\end{cases}
\end{equation*}
where $n_L$ is the number of ligand atoms, $n_P$ the number of pocket elements (atoms or vertices), and $d^*$ a distance threshold. For a visual representation of LC score, see Figure \ref{fig:evaluation}.
A high Ligand Coverage score denotes a pocket in contact with most of the co-crystallized ligand. However, this score alone does not exclude very large surface regions that go beyond the real pocket. Thus, limiting the evaluation of a prediction to the LC score would be a poor estimate of the quality of a prediction (virtually, the whole protein structure would score 100$\%$ on LC score).

\paragraph{Pocket Coverage score} It represents the fraction of the surface belonging to a pocket which is within a threshold distance from any ligand heavy atom. This score is very similar to the former, but it is referred to the pocket (the normalization is thus given by the total number of atoms or vertices constituting the putative pocket):
\begin{equation*}
	\mathrm{PC} = \frac{1}{n_{P}}\sum_{i=1}^{n_P}\delta_{ij}
	\quad\text{for }\forall\, j\in \mathcal{L}\,\,\text{with }\delta_{ij} =
	\begin{cases}
		1\,\,\text{if } d(i,j)\leq d^*\\
		0 \,\,\text{if } d(i,j)> d^*
	\end{cases}
\end{equation*}
For a visual representation of PC score, see Figure \ref{fig:evaluation}.
A high Pocket Coverage score implies that the putative pocket is mostly in close proximity of the ligand. Again, this score alone would not be sufficient to evaluate correctly a prediction. Indeed, too small pockets with respect to a larger co-crystallized ligand would score very high in PC but could be missing large parts of the binding region.

\paragraph{Threshold values}
For pockets expressed in terms of protein atoms (PQR files) we use $d^*=5$ \AA.
Since triangulation vertices are closer to the solvent than atom centers, we adopt a smaller distance threshold $d^*=4$ \AA when evaluating putative pockets based on OFF files.
These choices reflect the thresholds used when generating the training set.

Finally, a putative pocket is considered to be a correct match if it scores at least $50\%$ in Ligand Coverage and at least $20\%$ in Pocket Coverage.
When evaluating the results, we keep track of these scores separately, so as to have a more detailed statistical characterization of the performance of a method.

\section{Description of methods}
\label{sec:methods}

Eight groups from four different countries registered to this track. Four of them proceeded with the submission of their results.
In the following, we denote the methods proposed by the four participants as M1, M2, M3, and M4.

Specifically,
\begin{itemize}
\item method M1 is proposed by Hao Huang, Boulbaba Ben Amor, and Yi Fang;
\item method M2 is proposed by Yuanyuan Zhang, Xiao Wang, Charles Christoffer, and Daisuke Kihara;
\item method M3 is proposed by Apostolos Axenopoulos, Stelios Mylonas, and Petros Daras;
\item method M4 is proposed by Luca Gagliardi and Walter Rocchia.
\end{itemize}
Lastly, Luca Gagliardi, Andrea Raffo, Ulderico Fugacci, Silvia Biasotti, and  Walter Rocchia are the organizers of the SHREC 2022 track on protein-ligand binding site recognition.

The remaining part of this section is devoted to describe the 4 proposed methods.
While each method will be discussed in detail in the corresponding subsections, one could preliminary classify them based on their input format and on their adopted strategy. Concerning the input, M1 is the only method adopting the OFF files as a representation of the molecular surface, while M2, M3, and M4 feed their approaches with models expressed through anonymized PQR files.
Differently, focusing on the proposed strategies, M1, M2, and M3 exploit statistical learning, while M4 adopts a direct approach.

\subsection{M1: Point Transformer}
\label{subsec:M1}

\begin{figure}[htb!]
\centering
  \includegraphics[width=0.85\textwidth]{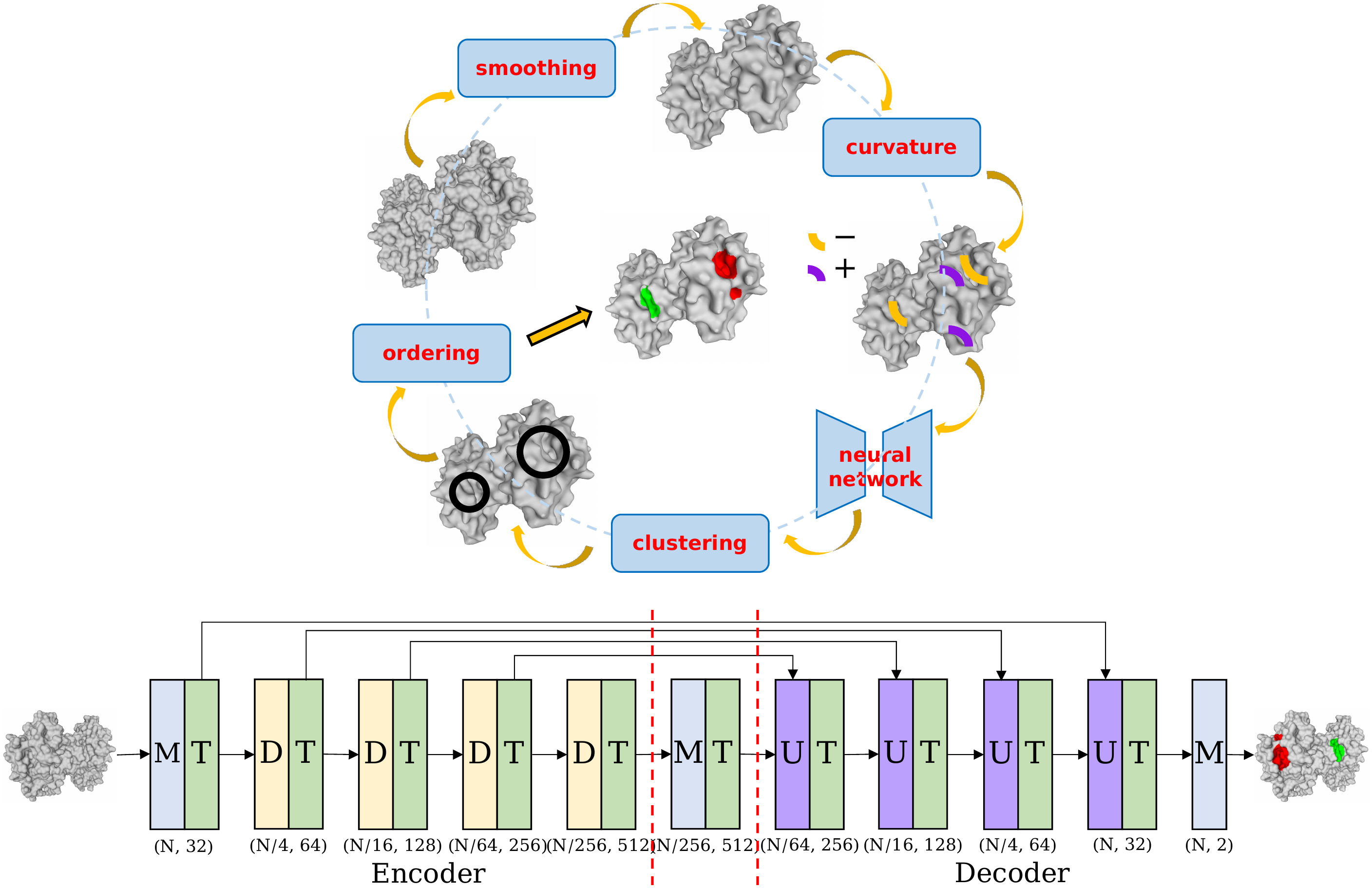}
  \caption{A graphical representation of the strategy adopted in method M1. The figure on the top displays the pipeline while the one on the bottom depicts the model architecture. ``-'' sign (orange) and ``+'' sign (purple) represent negative and positive curvature values per-vertex. The label ``M'' stands for multi-layer perceptron (MLP), ``T'' for point transformer layer, ``D'' for transition down, and ``U'' for transition up. The bottom numbers indicate the number of vertices and channel size.}
\label{fig:M1}
\end{figure}

Transformer was originally proposed for machine translation and it has achieved notable performance on various computer vision tasks  \cite{Transformer1}. Due to the fact that input proteins are provided in the form of triangulated meshes, a Transformer-based neural network model \cite{Transformer3} is adapted to learn per-vertex local shape geometric features. The Transformer-based model initially developed for the purpose of segmenting 3D point clouds is customized to segment binding regions on 3D protein shapes. After being trained on a relevant dataset of protein shapes, the model is able to learn discriminative per-vertex local shape descriptors for binding region prediction.
A visual description of the pipeline adopted in method M1 is depicted in Figure \ref{fig:M1}. Starting from a given protein mesh (top-left), mesh surface is smoothed using Laplacian, then per-vertex curvature is computed. A 5-dimensional vertex feature (i.e., coordinates and curvatures) is fed into a Transformer-based neural network to predict a binary segmentation result as a ligandability score. Finally, the candidate binding region vertices are clustered based on their ligandability scores and the binding regions are formed and ordered according to the vertex scores within each region.

\subsubsection{Feature extraction}
The first step of the proposed method smooths mesh surfaces by adjusting vertex positions using Laplacian smoothing  \cite{Transformer2}, with the effect of ``relaxing'' the meshes, making the triangles better shaped and the vertices more evenly distributed. Specifically, for each vertex $v$, a list of vertices $N(v)$ which are directly connected to $v$ is determined. Then, an iteration phase begins over all vertices. For each vertex $v$, the coordinates of $v$ are updated according to an average of the connected vertices $N(v)$. A relaxation factor $r$ is applied to control the amount of displacement of $v$. The process repeats $n$ times for each vertex until the desired result is obtained. In the experiments, the parameters $r$ and $n$ are set as $0.2$ and $200$, respectively. For the smoothed meshes, each vertex $v$ is represented by a 5-dimensional feature vector $f_v = [x, y, z, g, m]$ where $[x, y, z]$ denotes the normalized Euclidean coordinates and $[g, m]$ denotes Gaussian curvature and Mean curvature, respectively. The feature vector $f_v$ is then fed forward to a neural network model as described below.

\subsubsection{Adopted neural network}
Point Transformer  \cite{Transformer3}, a Transformer-based neural network model as shown on the bottom in Figure \ref{fig:M1}, is employed as it has achieved state-of-the-art performance on point cloud object shape classification, shape part segmentation and scene segmentation. The network adopts a U-Net  \cite{Transformer4} architecture consisting of an encoder and a decoder. The encoder consists of five blocks and each block contains a transition-down layer to reduce shape resolution and a point transformer layer to aggregate local geometric features for each vertex, except for the first block containing an multi-layer perceptron (MLP) layer to expand each vertex feature from 5 dimensions to higher dimensions. Similarly, the decoder consists of four blocks and each block contains a transition-up layer to recover shape resolution and a point transformer layer serving the same purpose as in the encoder. The tailing MLP layer is utilized to regress the final results.
We refer the reader to  \cite{Transformer3} for the description of internal layer structures.

The binding region prediction is treated as a binary shape segmentation where $1$ represents the class of binding regions and 0 denotes the class of non-binding regions. A weighted cross-entropy loss is employed to train the network. The weight for each class in the loss is inversely proportional to the number of vertices belonging to the corresponding class. The proposed model is trained on the provided dataset, which comprises 935 protein shapes for training and 165 shapes held out for testing.

\subsubsection{From scores to binding sites}
For each vertex $v$, a ligandability score is defined as $LS(v):=\max\{p_1^v - p_0^v, 0\}$ where $p_i^v$ is the un-normalized probability generated by the network for class $i$. Ligandability is intended as the capability of a given region to bind a ligand, not necessarily resulting in a biological outcome \cite{surade2012structural}. To prepare putative binding region predictions, the vertices that have ligandability score lower than a given threshold (default $t = 2.0$) are filtered out. Then, the remaining vertices are candidate to form bind regions and we need to group them spatially. Due to the fact that different protein meshes have varying numbers of binding regions which are unknown in advance, we cannot utilize K-means or hierarchical clustering algorithms. We instead opt to cluster using the DBSCAN  \cite{Transformer5} algorithm. We input per-vertex ligandability scores (above the threshold) to DBSCAN and the two parameters $eps$ and $min\_samples$ in DBSCAN are set to the average edge length and 5, respectively. Predicted binding region is then formed by the set of vertices in a cluster. Next, similar to  \cite{Transformer6}, each region is assigned a score calculated as the average of squared ligandability scores of all of the vertices that define the region:
$$
BRScore := \frac{1}{|C|} \sum_{i=1}^{|C|}LS(v_i)^2,
$$
where $|C|$ is the number of vertices in a cluster. Squaring of the ligandability scores puts more emphasis on the vertices with higher ligandability score (i.e., vertices that are classified as ligandable with more confidence). The very last step involves reordering the putative binding regions in a decreasing order of their $BRScore$ and assigning positive integer ranks to each binding region with the most confident region assigned with the smallest rank number.

\subsubsection{Computational aspects}
The experiments are performed on a machine with an Intel(R) Xeon(R) E5-2680 v4 2.40GHz CPU supporting AVX2 and two GPUs V100 with 32GB of memory each, with 80GB RAM memory.
The code for vertex feature extraction uses the APIs from Visualization Toolkit (VTK), the implementation of the network is partially adapted from Point Transformer  \cite{Transformer3} written in Python with PyTorch 1.7.1 as the deep learning library, and the DBSCAN utilizes the API from scikit-learn.
The training takes around 15.5 hours with a batch size of 8 for 100 epochs.
The time required for feature extraction (smoothing and computing curvature) is around 40 minutes for the training set and around 5 minutes for the testing set.
The computation of the score for the test set through the trained network takes around 2 minutes for the testing set.
The time for identifying the binding sites from the score is around 1.5 minutes for the testing set.

\subsection{M2: GNN-Pocket}
\label{subsec:M2}

Method M2, named GNN-Pocket, is developed to detect pockets on protein surfaces and it is based on the use of a graph neural network (GNN). VisGrid  \cite{GNN1} and ghecom  \cite{GNN2} is adopted to extract features for each atom. Then, a graph with these atoms is constructed. Finally, a 4-layer GNN  \cite{GNN3, GNN4} is developed in order to return, for each atom in the input protein surface, its probability of belonging to a pocket.

\begin{figure}[htb!]
\centering
  \includegraphics[width=0.85\textwidth]{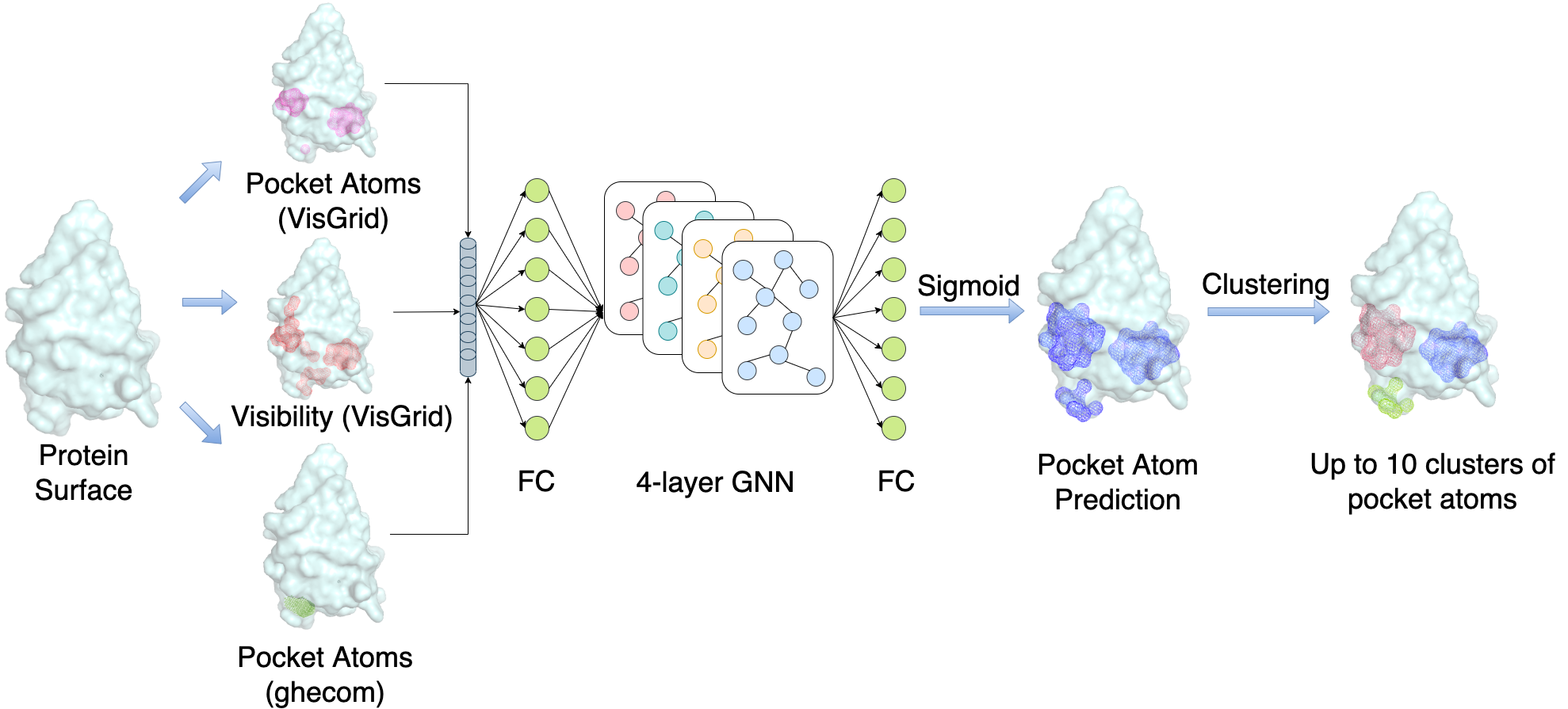}
  \caption{A graphical representation of the strategy adopted in method M2.}\label{fig:M2}
\end{figure}

A visual description of the pipeline adopted in method M2 is depicted in Figure \ref{fig:M2}.

\subsubsection{Feature extraction}
First, three types of features are collected by VisGrid and ghecom. VisGrid  \cite{GNN1} uses a voxel-based visibility criterion to identify pockets in a protein structure. Ghecom  \cite{GNN2} identifies deep and shallow pockets of using spherical probes of different sizes. The first feature is a binary output from VisGrid, which indicates if an atom has a visibility lower than a cutoff. The second feature is the number of closest grid points that are predicted as pockets by ghecom. As the third feature, the number of grid points within 8\AA that are predicted as pockets by VisGrid is chosen. Finally, three features are concatenated into a 3-dimensional vector as the input embedding of the neural network.

\subsubsection{Adopted neural network}
The collected features are adopted for constructing graphs having as nodes the atoms of the input protein. After having evaluated about a dozen GNN models with different feature combinations and graph choices, two graphs are constructed using different edge connecting criteria. In $Graph 1$, an edge is built if the distance between two atoms is smaller than the sum of their radius. In $Graph 2$, two atoms are connected if they are closer than the sum of radius plus the size of a water molecule (2.8 \AA).
To train the GNN models, the provided dataset of $925$ proteins is split into two sets: $740$ proteins for training and $185$ for validation. In training, Dice loss, which considers the intersection and union of a prediction and the ground truth, is adopted. The method makes use of the Adam optimizer with the following configuration: a learning rate of $0.001$; a linear learning rate decay; an exponential decay rate for the $1^{st}$ momentum estimate of $0.9$; an exponential decay rate for the $2^{nd}$ momentum estimate of  $0.999$; a weight decay of $1e-6$.

\subsubsection{From scores to binding sites}
Among the constructed GNN models, four models that have a relatively high recall or F1 score are chosen. Then, an ensemble model that combines the four models by averaging their pocket probabilities is considered. Atoms are predicted as within a pocket region if their probability is higher than $0.5$.
Since it shows the highest F1 score, the ensemble model is selected as final model. In the test set provided by the organizers, there are $5$ proteins where prediction does not include more than $10$ pocket atoms. In those cases, VisGrid output is directly used as prediction.
A bottom-up hierarchical clustering method, which minimizes the distance between the closest pairs of clusters, is adopted to group pocket atoms into pocket regions. The top-10 pockets by the sum of probability values of atoms are selected.

\subsubsection{Computational aspects}
The GNN model is trained on a machine with an Intel(R) Xeon(R) 3.60GHz CPU and a NVIDIA RTX 2080Ti GPU, with disk memory of 3.7 TB. The language for model implementation is Python. For training stage, each GNN model takes around 24.5 hours with 100 epochs. For inference stage, it takes 1 minute 42 seconds for a structure of 2269 atoms. Feature extraction takes 1 minute 30 seconds, including 5 seconds to run VisGrid, 7 seconds to run ghecom, 1 minute 18 seconds to build adjacency matrix and prepare feature embedding. GNN model takes 12 seconds to do inference. For model ensemble and clustering stage, it takes 2 seconds to ensemble predictions and 10 seconds to get clustered pockets.

\subsection{M3: DeepSurf}
\label{subsec:M3}

\begin{figure}[htb!]
\centering
\includegraphics[width=0.85\textwidth]{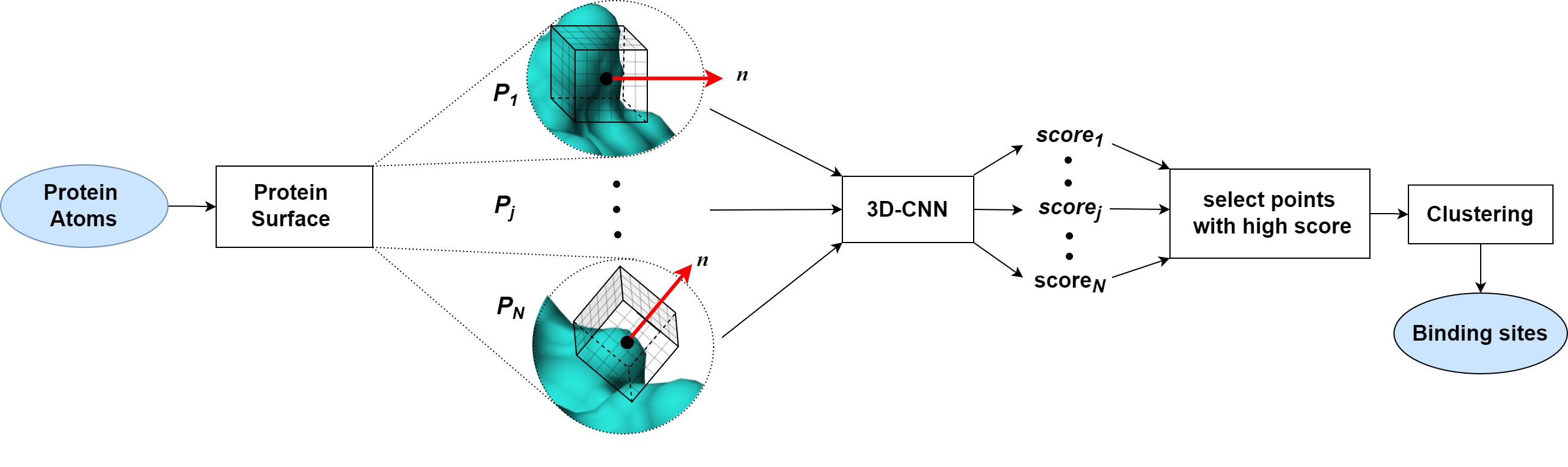}
\caption{A graphical representation of the strategy adopted in method M3.}\label{fig:M3}
\end{figure}

The strategy adopted by M3 follows the recent advances in the machine learning field and the
extensive application of deep learning methods on various tasks. More specifically, M3 employs
DeepSurf \cite{Deep1}, a recently proposed deep learning approach for the prediction of potential binding sites on proteins. DeepSurf combines state-of-the-art deep learning architectures with a
surface-based representation, where a number of local 3D voxelized grids are placed on the
protein surface. A visual description of the pipeline adopted in method M3 is depicted in Figure \ref{fig:M3}.

\subsubsection{Feature extraction}

Firstly, if not provided, the molecular surface of the protein is created in a triangular mesh format.
Then, an optional mesh simplification step takes place to avoid unnecessary redundancy of points. This is achieved by grouping adjacent surface points to clusters using the K-means clustering algorithm, while keeping as representative point for each cluster only the closest one to the cluster center. The density of the remaining surface points is controlled by the parameter $f$, so as, if the initial surface points are $n_p$, the final ones are $n_p/f$. Centered at each sample point $P$ of the surface, a local voxel grid of size $16 \times 16 \times 16$ and
resolution 1\AA is computed. To achieve rotation invariance before the feature computation,
the local grid is oriented such that the $z$-axis is always parallel to the normal vector $n$ on $P$,
i.e., perpendicular to the surface (see Figure \ref{fig:M3}). The next step is to calculate the
necessary features for each voxel of the local grid, using the featurization scheme proposed originally in \cite{Deep2}. According to this scheme, 18 chemical features are calculated per protein atom and each grid voxel receives the features of the atoms inside it. This step requires information of the atom types in order to calculate the necessary features. Since in this track, the provided protein files lack such information, this information was tried to infer from the atom radii, which in general can be regarded as a highly confident indication of the atom type.

\subsubsection{Adopted neural network}

The previously described steps allow to form a 4D tensor for each surface point $P$, which is imported to a 3D-CNN and produces at the output a ligandability score in the range of $[0,1]$. This score denotes the probability for the surface point $P$ of belonging to a binding site. The proposed methodology is generic, meaning that any 3D-CNN architecture that receives as input a 4D tensor and returns as output a float value in range $[0,1]$ can be used. Nevertheless, the considered architectures are two: a 3D-ResNet and a custom Bottleneck-3D-LDS-ResNet, which has shown in previous experiments similar performance with much fewer parameters \cite{Deep1}.

\subsubsection{From scores to binding sites}

After obtaining ligandability scores for all surface points, those points with score less than a
ligandability threshold $T$ are considered not reliable and are discarded, while the remaining
ones are clustered in the 3D space using the mean-shift algorithm. The created surface clusters correspond to the binding sites, which are sorted based on the average ligandability scores of their member points.

Finally, the surface points from each cluster are mapped to their closest protein atoms in order to obtain binding sites on the atom level.

\subsubsection{Computational aspects}

DeepSurf is originally trained on the large scPDB database \cite{Deep3}, which comprises $16034$
entries corresponding to $4782$ proteins with $17594$ total binding samples. For the needs of
the track, the original trained models are kept and experimented mainly on its hyperparameters
using as validation set the training set provided by the organizers. As a result of this
experimentation, it is decided to keep a dense surface point grid, (parameter $f$ is set to $1$,
meaning no simplification at all), while the ligandability threshold $T$ is set to $0.9$. Between the
two architectures, the computational heavier 3D-ResNet, showing a performance boost on the
validation set, is selected. The experiments run on a machine with a GeForce GTX1070 GPU and an Intel(R) Core i7- 6700K CPU. The method is implemented in Python and uses the Tensorflow deep learning framework. The training takes 17 hours on the GPU. The inference time to extract the predicted binding sites on the test set is 3.6 hours, utilizing both GPU and CPU.

\begin{figure}[htb!]
\centering
\includegraphics[width=0.6\textwidth]{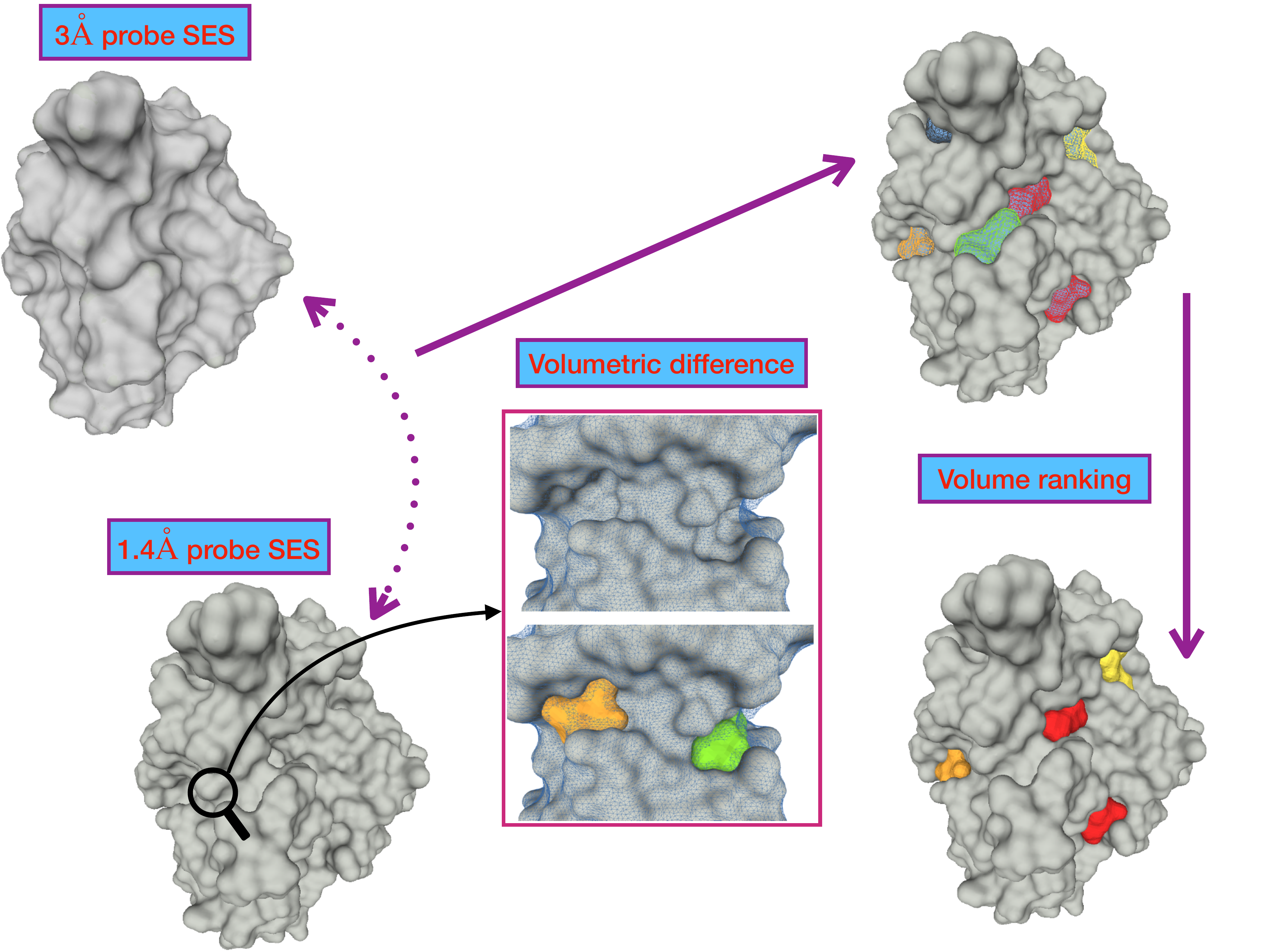}
  \caption{A graphical representation of the strategy adopted in method M4. NanoShaper creates two SES molecular surfaces at probe radius 1.4\AA (standard average water molecule radius) and 3\AA. Pockets are defined as the enclosed cavities between the two meshes (central panel, light blue 3\AA triangulation vertices). Pocket surfaces are constructed by filling the identified cavities with water spheres.
  For illustrative purposes, the three largest pocket detected (Top3 ranked) are represented by the red, orange and yellow meshes, in decreasing order.
  Note that here the largest pocket is a tunnel.
  }\label{fig:M4}
\end{figure}

\subsection{M4: NS-Volume}
\label{subsec:M4}

NanoShaper (NS) is an efficient software for triangulating complex manifold surfaces based on an ad-hoc ray-casting approach and the CGAL library \cite{DeCherchi2013}. NS can build molecular surfaces according to several definitions: skin, blobby, and the SES (Connolly) molecular surfaces.
Geometrical patches are first calculated analytically and an accurate triangulation (\emph{Marching Cube algorithm}) is drawn from the analytical intersections of these patches with grid-rays. In the process, the volume and surface area is also calculated.
In this SHREC track NS is first of all used to generate the Solvent Excluded Surface (SES) of the dataset. In NanoShaper, the SES is built according to alpha shapes theory which allows the derivation of accurate analytical geometrical patches \cite{Chen2010}.
Even if NS was mainly designed for the triangulation of molecular surfaces, it offers also a pocket detection function.
Pockets are defined as the volumetric difference between the space regions
enclosed within the SESs of the protein obtained with two
different probe radii, 1.4 \AA (water molecule effective radius) and 3 \AA.
The implementation is grid based by flagging those grid points which are simultaneously inside the $3$ \AA SES and outside the $1.4$ \AA SES.
Once the grid points are identified, a filtering procedure is adopted which preserves points which are i) within $1.4$ \AA from all flagged point or ii) within $1.4$ \AA from points fulfilling i). Pockets are then defined as the unconnected components on the grid after the filtering by applying a \emph{flood-fill} procedure \cite{Decherchi2019}. Then the pocket surface is constructed by building the molecular surface of the union of water spheres ($1.4$ \AA) centered on the pocket grid points.
Only pockets above a threshold of three water molecules volume are returned.
By default, the pockets returned by NS do not follow any specific ordering.
Here we implement on top of NS a simple sorting of the pockets by volume, from the largest to the smallest. The output provided is for each pocket  a surface mesh (OFF format), and a list of atoms contacted by the pocket (a subset of the whole protein in PQR format). The overall pipeline is illustrated in Figure \ref{fig:M4}.
This methods does not contain any learning. It is here proposed for comparison against the data-driven methods discussed above, showing the effectiveness of simple volume ranking as a strategy for ligand binding site recognition.

\subsubsection{Computational aspects}
Pocket detection speed depends on the size of the structure, since it is based on the construction of two SES surfaces over the whole protein. In general, NS has proven to be very efficient in comparison to most molecular surface construction softwares \cite{DeCherchi2013}. Furthermore, NS can be speed-up with multi-threading (reported 10X speed up on 8-core machine with respect to single core) \cite{DeCherchi2013}.
We write a custom python script which calls externally NS using the ``pocket'' function, extracts information on the putative pockets volume, and ranks them accordingly. As mentioned previously, we also extract information on the protein atoms contacted by the pockets. This can be used to construct a labelled PQR with the same format as the one given for training to be used in this contest.
The algorithm runs on a Intel(R) Core i7-1085H CPU (2.7 GHz).
On the test set using single core, the average execution time per structure is of about $5\,$s and we measured a maximum of about $16\,$s for the largest structure (PQR code 5ykw).





\section{Comparative analysis}
\label{sec:results}

\subsection{Ranking protocol}
The performance of each method presented in Section \ref{sec:methods} are here quantitatively evaluated. As described in Section \ref{sec:eval_meas}, a putative pocket, in the form of a list of labelled OFF vertices or PQR atoms, is considered a correct match if its Ligand Coverage (LC) score is above $50\%$ and Pocket Coverage (PC) score above $20\%$. Results for a method are then summarized evaluating the effectiveness of the returned ranking in terms of average successfully predicted pockets.
Similarly to what proposed in Refs. \cite{Krivak2018,Mylonas2021}, for a given structure with one or multiple known co-crystallized ligands and a method returning a ranked list of putative pockets, the ranking position is given by the number of preceding non-matching pockets. The normalization is given by the number of structure-ligand pairs.
In this manner, we ensure that results are comparable across structures with differing number of pockets observed binding a ligand. For instance, if a structure has four known binding sites and these are all matched by the first four ranked pockets, this would be considered as $100\%$ Top1.
\subsection{Results}


\begin{table*}[ht]
	\centering
\begin{tabular}{l|c|c|c|c|c|c}
	\toprule
Method  & Top1   & Top3   & Top10  & LC & PC & nPockets \\
\midrule
M1 - Point Transformer & $69.1$  & $75.9$ & $75.9$ &   $96.4$   &  $60.4$     & $2.1$  \\ 
M2 - GNN-Pocket & $53.4$  & $54.6$ & $55.4$ &    $93.7$   &  $47.5$     & $1.9$   \\ 
M3 - DeepSurf & $87.6$ & $89.2$ & $89.2$ & $95.0$ & $67.9$ & $1.6$ \\ 
M4 - NS-Volume & $59.0$ & $76.7$ & $83.9$ & $88.8$ & $74.8$ & $11.6$\\ 
\midrule
Fpocket\textsuperscript{\emph{a}} & $60.2$ & $75.1$ & $84.7$ & $92.5$ & $64.7$ & $8.9$\\
\bottomrule
\end{tabular}

{\textsuperscript{\emph{a}} Standard pocket detection method  \cite{LeGuilloux2009}, for comparison purposes.}
	\caption{Results are expressed as a percentage representing the success rate normalized over the total number of structure-ligand pairs.
    We also report on the average LC and PC scores of successful pockets and average number of putative pockets per structure generated by each method.}\label{tab:results}
\end{table*}

\begin{figure}
    \centering
    \includegraphics[width=0.7\linewidth]{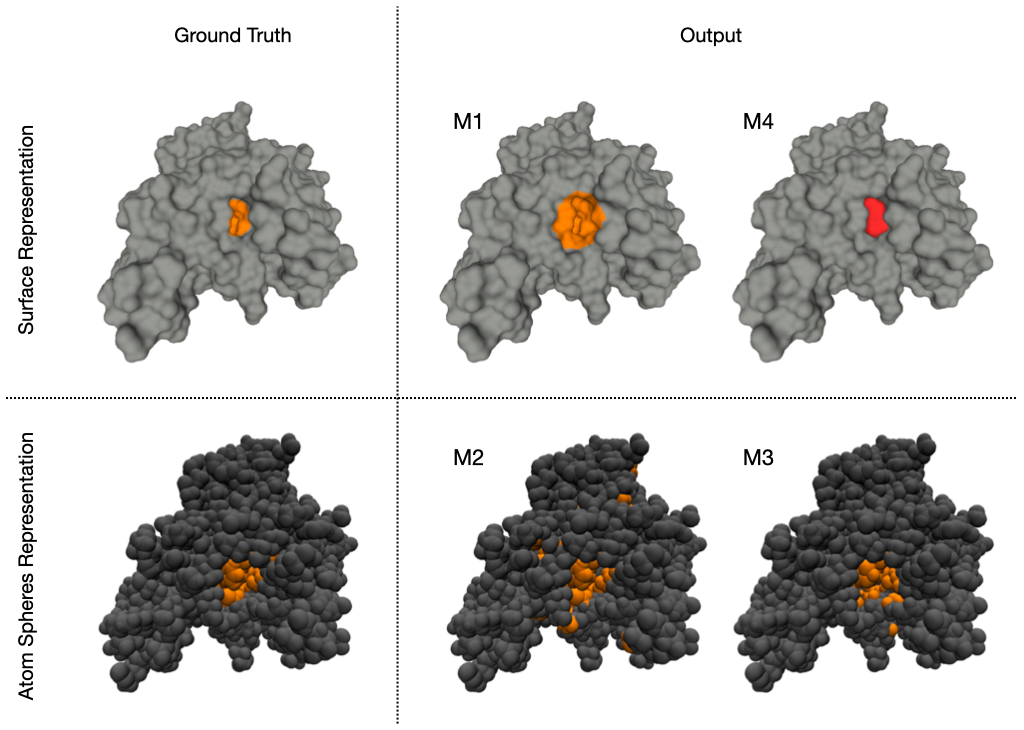}
    \caption{
    An example of successfully predicted pockets on the structure with PQR code 3lcv.
    In the left column, the ground truth ($LC=100\%$ and $PC=100\%$) of the considered structure, in the right one, the output returned by the proposed methods (predicted sites are highlighted in orange).
    The output of the various methods is displayed in accordance with the representation they adopt (mesh surface in the top row, atom spheres in the bottom one).
    M1-Point Transformer: pocket evaluation measures are $LC=100\%$ and $PC=76.5\%$.
    M4-NS-Volume: Only the largest pocket is shown.
    Protein atoms contacted by the pocket mesh are used for scoring, $LC=100\%$ and $PC=88.6\%$.
    M2-GNN-Pocket: $LC=100\%$ and $PC=36.2\%$ (not visible: orange spheres on the opposite side).
    M3-DeepSurf: $LC=100\%$ and $PC=87.4\%$.
    }\label{fig:discussion}
\end{figure}

The performances on the test set of the different methods are summarized in Table \ref{tab:results}. In addition to evaluating the average ranking performance in terms of Top1, Top3, and Top10 (maximum allowed number of pockets returned) performance, we report on the average LC and PC scores over successfully predicted pockets, and average number of generated pockets per structure.
Each line refers to one of the methods analysed. For sake of comparison, we added an extra last line describing the results obtained by Fpocket on the same dataset.
Fpocket is a standard and well established tool for pocket detection \cite{LeGuilloux2009}. This method uses as input PDB files (containing full chemical information) and so it would not be appropriate to this SHREC track which focuses on geometric rather than chemical features.

As a general comment, we notice that only M4 and Fpocket return more than about 2 putative pockets per structure on average. If on the one hand this could be appreciated by the user, since it provides a more concise information, on the other hand, it is potentially detrimental since it reduces the probability to find more reasonable candidates and to find more binding pockets in a single protein.
The method which gives the best results, outperforming significantly also Fpocket, is M3 - DeepSurf. Indeed, despite the small number of putative pockets generated, these are extremely well predicted, ranking 89$\%$ in Top3 ($10\%$ higher than Top3 of Fpocket). It is interesting to note that overall, Fpocket and M4 - NS-volume score very similarly to M3 on Top10, showing the outstanding capacity of M3 to pinpoint the observed binding pockets within the top ranked, but eventually not surpassing significantly the other methods on the total number of successful predictions generated.
It is important to note, however, that the procedure adopted by M3 is partially beyond the scope of this work. Indeed, as described in section \ref{subsec:M3}, M3 - DeepSurf is based on 18 \emph{chemical} features that are imported in a CNN. The chemical features, in this case the atoms, can be deduced from the information on the atomic radius provided in the anonymized PQR. Furthermore, the CNN was previously trained on a distinct dataset (the training set here proposed is used only for hyper-parameters optimization).  This somehow reduces the possibility to use M3 to really assess the power of purely geometrical methods in both generating and ranking the putative sites. Second, even assuming that the atomic radius can be considered on par with a label, any training should be restricted to the given training set, in order to evaluate all the methods on the same footing. Finally, we are not aware about the degree of overlap between our test set and the large dataset used to originally train M3. In any case, the results of M3 remain certainly impressive and a successful example of \emph{transfer learning}.
It is also interesting to note that M4 (i.e., NS-Volume) is very competitive even if it is not based on Machine Learning: the pockets are generated by a purely agnostic geometrical method, and then simply ranked by volume.
M1 is the only method based on OFF files. It is worth noting that its Top1 performance is extremely good, slightly outperforming Fpocket. However, similarly to M2 and M3, the amount of successfully predicted pockets does not increase importantly when considering the next ranked Top3. This behavior is a consequence of the low number of returned pockets.  M4 - NS-Volume and Fpocket, which return on average about 10 pockets, show about a $20\%$ and a $14\%$ increase in successful prediction when moving from Top1 to Top3, respectively.

Moving to the quality of the correctly detected pockets, we observe that all methods perform very well in term of Ligand Coverage score, while a significantly lower Pocket Coverage score is measured. A low average PC score indicates that a method is prone to generate pockets which are larger than the binding ligand. To illustrate qualitatively the significance of PC, in Figure \ref{fig:discussion} we show as an example a pocket correctly identified by all methods. It can be observed, that the binding pocket of M2 is overestimating the actual binding region of the ligand.
In general, we observe that M2 generates pockets which are often larger than the binding region and scattered into disconnected segments. This type of behaviour could be due to the ``translation cost'' required for applying a voxel-based representation to a model expressed in the form of a PQR file, or to the lack of further post-processing.
The systematic generation of excessively large pockets in M2 is also statistically reflected in Table \ref{tab:results} by the lower average PC score with respect to the other methods.

\begin{figure}
    \centering
    \includegraphics[width=0.4\linewidth]{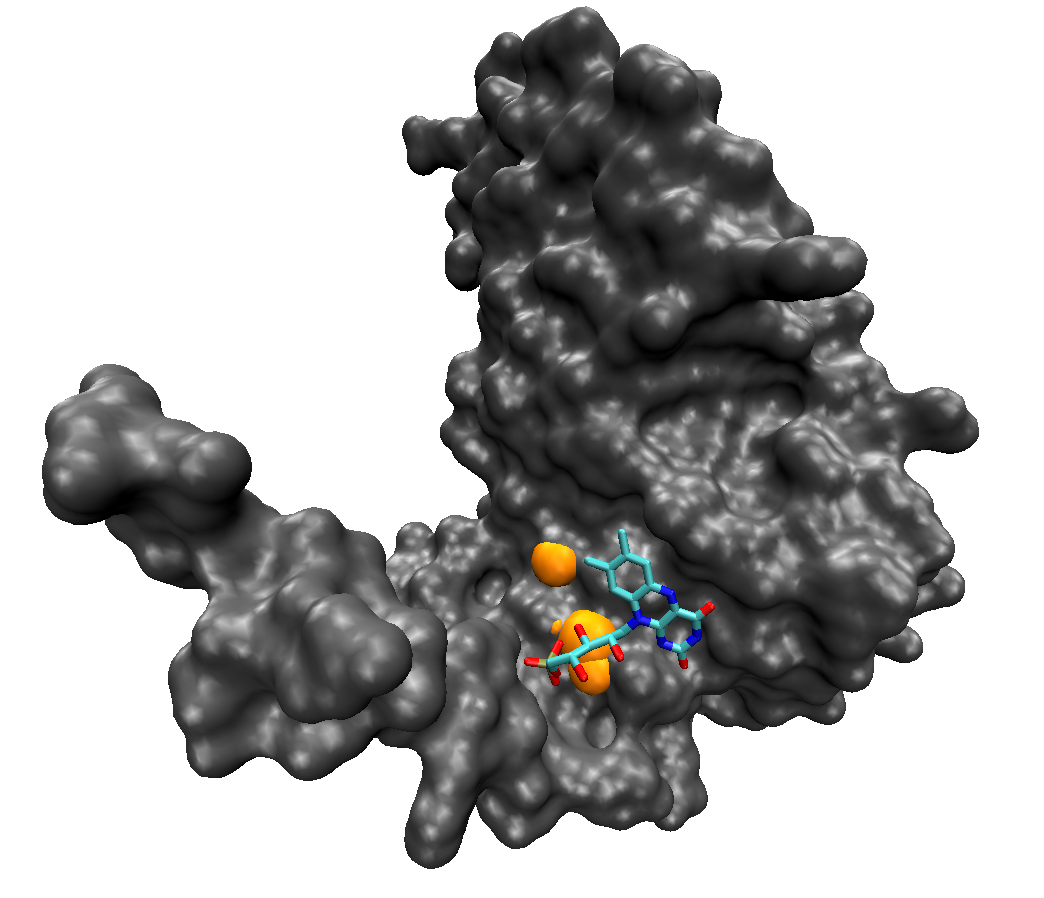}
    \caption{An example of a shallow binding site on the structure with PDB code 1nox, which cannot be correctly identified by any of the methods. The ligand is also represented. In orange, the protein atom spheres of the pocket predicted by M3 - DeepSurf (artificially enlarged for visualization purposes). This pocket is not considered a match since too small (below the LC threshold).}
    \label{fig:comparison}
\end{figure}

As often observed in other pocket detection algorithms \cite{Simoes2017}, the proposed methods have more difficulty in identifying particularly shallow binding sites.
Indeed, methods completely relying on geometry for the generation of putative pockets are optimized to recognize cleft and cavities (which in fact are often found to contain binding ligands). Differently, due to their anomalous geometric nature, shallow pockets are more difficulty retrieved since they attain the role of binding site mainly for their chemical properties rather than their shape.
As an example, Figure \ref{fig:comparison} depicts a structure having a shallow binding site which is not identified by any of the proposed methods.
Methods as M1, M2 and M4 (as well as Fpocket) fail the detection because they do not adopt chemistry-related information. This limitation is not fixed by adopting a learning-based approach, as for M1 and M2, and is probably due to the fact that shallow sites are rare in the training set.
However, from those ML methods such as M3 - DeepSurf, which leverage also on chemical features to form putative sites, one would expect the ability to highlight locations of high protein-ligand chemical affinity, regardless of the geometry. As depicted in Figure \ref{fig:comparison}, M3 - DeepSurf nevertheless also seems to deteriorate when trying to identify shallow sites. In this specific example, only a few (too little in order to form a relevant pocket) number of atoms are identified in the region of interest by DeepSurf.
We think that this behavior is attributed to the high selected value of the ligandability threshold T (T=0.9) which was tuned on the training set. Selecting a lower value (e.g., T=0.5) could make the algorithm more aggressive and lead to the selection of more surface points and, consequently, to the formation of larger binding sites. On the other hand, a lower ligandability threshold could lead to a larger number of false positives. As expected, different types of sites might require different hyperparameter values.

\section{Concluding remarks}
\label{sec:obs_conlusion}

In this paper, we provide a detailed analysis and evaluation of four algorithms dealing with the problem of automatic detection of binding sites on protein surfaces (aka ligandable pockets) given a training database of positive examples (no negative labelled data). The database was presented as a molecular surface mesh (OFF format) or a PQR file where the vertices or, respectively, the atom spheres are labelled according to whether they belong or not to a known ligandable pocket.

The performance of each method is evaluated in terms of two measures: the Ligand Coverage score (LC), and the Pocket Coverage (PC) score. Together they express the ability of a binding pocket predictor to find the smallest region which binds a ligand on the protein molecular surface. Most of the proposed methods showed very good performance, comparable to that of Fpocket, a pocket detection tool widely used in the Computational Biology community. In particular, method M3 showed an outstanding performance. However, we observed that the information leveraged by M3 goes beyond that of pure geometry and the training set is larger than the one provided to the participants to the track.
Interestingly, a method based solely on a purely geometric technique and no learning, M4, confirms that simple volume based scoring on geometrically identified pockets remains a very valid approach. This type of ranking is actually adopted by several standard geometry based pocket detection algorithms \cite{Laskowski1995,Hendlich1997,Weisel2007,Tripathi2010,Tian2018}. We observe that all methods perform worse on shallow binding sites. This is a common limitation in pocket detection softwares mainly due to the fact that binding ligands are mostly found in deep clefts and grooves.
Finally, we observed that it is generally hard to perform high on Pocket Coverage.
Given a high LC score, a low PC score is related to a pocket which is exceedingly large with respect to the ligand binding region.
Therefore, this suggests a margin for improvement in the direction of a higher segmentation of the returned sites into separate smaller pockets
or sub-units (sub-pockets \cite{Marchand2021,Volkamer2012}). Given the effectiveness of the ML approaches proposed, such a segmentation might be able to identify more precisely the exact binding site without compromising unreasonably the effectiveness of the ranking.

The problem proposed in this SHREC track is an instance of a one-class discrimination task, since experiments can only provide positive examples. However, some of the methods discussed in this work, which are based on a learning process, turn the problem into a two-class discrimination task by labelling as negative the surface regions or atoms which do not belong to experimentally observed binding sites. This points to a critical aspect of this task, namely whether it can be effectively mapped into a standard ML problem, especially when employing DNN and large datasets (as done by M3). Further studies are needed to assess the precise nature of the boundary between the conceptual nature of the problem and practical applications.

The benchmark, as well as the participants' predictions that originated the results described in Section \ref{sec:results} and in the appendices, are available at \url{https://github.com/concept-lab/shrec22_proteinLigandBenchmark}.

\section*{Online software repositories}
For the sake of replicability, for each of the four proposed methods we provide the link to the online software repository.
\begin{itemize}
    \item M1 - Point Transformer is available at
    \url{https://github.com/aaron-h-code/Protein_SHREC2022/},
    \item M2 - GNN-Pocket is available at \url{https://github.com/kiharalab/GNN_pocket},
    \item M3 - DeepSurf is available at \url{https://github.com/stemylonas/DeepSurf_SHREC22},
    \item M4 - NS-Volume is available at \url{https://github.com/concept-lab/NS\_pocket}.
\end{itemize}

\section*{Acknowledgements}

The track organisers thank Dr. Michela Spagnuolo for the fruitful discussions.
\\
The CNR-IMATI research is partially developed in the activities DIT.AD021.080.001 and DIT.AD021.125.
\\
The participants proposing method M2 thank Zicong Zhang, Yuki Kagaya, Jacob Verburgt, and Genki Terashi for technical help and discussion.
\\
This work is partly supported by fundings from the National Institutes of Health (R01GM133840, R01GM123055, and 3R01GM133840-02S1), the National Science Foundation (CMMI1825941, MCB1925643, and DBI2003635), and the Inception Institute of Artificial Intelligence (NYUAD Global Ph.D. Student Fellowship).

\bibliographystyle{ieeetr}

\end{document}